# A Tablet Based Learning Environment

Hong Cao

## Abstract

Pen computing tools such as Tablet PC, Tablet Monitors and its various supporting software tool offer another dimension to enhance our today's digitally integrated and connected classroom learning environment. This paper first reviews the various state-of-the-art pen-computing hardware and software that have been applied in the classroom setting to introduce student-centric learning, collaboration and making annotations and designing classroom activities easier. We then propose a new classroom environment which is fully equipped with Tablet devices and the supporting software tools for the goals of 1) easy electronic ink annotations with least constraints; 2) enhanced active learning with timely feedback; 3) enhanced student collaborations and 4) lecture recording. The classroom has been put into practical teaching and learning environment as a pilot project in our higher learning environment. After overcoming the initial learning curves, the environment received positive feedbacks from the teaching faculties as well as from the students.

## 1. Introduction

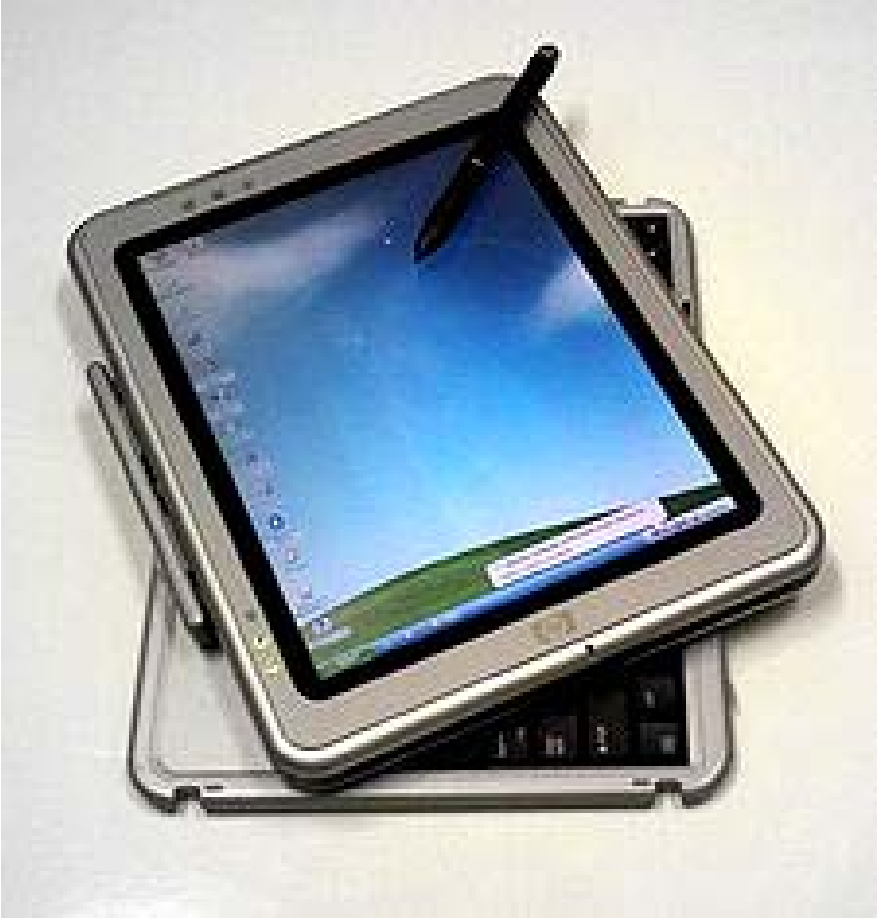

Fig. 1 A HP Compaq Tablet PC with Detachable Keyboard [1]

Since Bill Gates of Microsoft first demonstrated a Tablet PC prototype (See Fig. 1 as a Tablet PC illustration) in 2001, this new invention has attracted worldwide attention. Different from the traditional notebooks, a Tablet PC is equipped with a stylus pen and a screen digitizer, which allow people writing directly onto the screen. The handwritings are electronically captured, stored and displayed in the form of digital inks, which can be easily manipulated, re-organized, recognized, animated and transmitted over Internet via the computing power and today's digital technology. Though some people think Tablet PCs are merely special notebooks with pen features, many others recognize their huge potential as the revolution of PCs, which will change the way that human beings interact with computers. Over the past 8 years, the Tablet PCs have gradually become more popular. Most of the major PC manufacturers have developed their own commercial Tablet PC series and make them available to consumers. The Tablet hardware technology also gets better with a more affordable cost. As a key driver of Tablet PCs, Microsoft has not only built several Tablet PC operating systems (OS), but also developed electronic inking related APIs and many Tablet-based software applications. These APIs are made available to many other independent software vendors (ISV) to develop their own tablet PC software applications. Microsoft has also been actively involved in improving and promoting inking-related research, such as handwriting recognition, ink understanding, ink-searching and ink-animation. The outcomes of these research activities will make the inking feature to more useful to every Tablet PC users.

It is undoubtful that Tablet PCs will grow more popular in the future, especially when the cost of Tablet PC is further driven down, better Tablet hardware technology is developed, growing number of discipline-specific Tablet software applications are available and also after the TOUCH functionality is incorporated as in the recent Windows 7 and the latest Fujitsu Tablet PC models. All these features will make Tablet PCs more attractive to professionals in different industrial verticals. In health-care industry, some doctors in Singapore have already started writing their electronic ink prescriptions to promote automated processing. In art creation and digital media industry, high-quality tablet tools gradually become essential to improve the efficiency in handling multimedia files. In the education industry, Tablet PCs have long been viewed as an ideal tool, which combines the pen-writing capability, mobility and computing power, to support student-centric learning. As described in [4], “The Tablet PC has the potential to alter the educational process. This new technology significantly changes the way students and teachers interact. It adds completely new dimensions to classroom interaction by providing digital ink and drawing tools for writing, sketching, and drawing; and for real-time collaboration.” As pen-writing plays central role for classroom teaching and learning, this additional feature makes the instructor and students have more natural interaction and less waste of time on some necessary procedures, e.g. note taking.  One can refer to [4] for the comprehensive advantages and potential improvements one could expect from incorporation of Tablet PCs into the teaching and learning system. In the early study [2, 3] in Singapore’s two pilot schools, catholic high school (CHS) and crescent girl school (CGS), statistics have generally shown the positive altitude and perceptions among students and teachers of using Tablet-PC driven technology in classroom teaching. There are still many existing issues, such as the required IT proficiency of students and teachers, the imperfection of Tablet PC device and learning software tools, teacher’s experiences, attitude towards new technology and sharing of learning materials, the effects of Tablet PCs' introduction on low and medium achieving students and the distraction issue for students who cannot discipline them well in a learning process. Though the above issues can be important when Tablet PCs are first introduced, by making Tablet PC's based learning an inevitable mainstream and with better applications to foster the interaction between the instructor and the students and to control student's Tablet PC's usage, the comparative advantages of bringing the new technology is still salient. After recognizing the huge potential and backed by the encouraging statistics, more secondary schools have brought Tablet PCs into their daily teaching and learning process in Singapore.

For higher education, the merits and innovative uses of Tablet PCs and pen-based technology have been widely explored in Universities and tertiary institutions in the developed countries [5-29, 32-39]. Many ongoing research activities are being conducted on how to best use the Tablet PCs to improve the traditional educational pedagogy. In [16], J. Cromack explains how to Table PC embodies the 7 principles of learning-centric educational practices including

- Encourages student-faculty contact, i.e. fostering interaction;
- Encourages cooperation among students, i.e. sharing and collaboration;
- Encourages active learning attitude;
- Gives prompt feedback;
- Emphasizes time on task;
- Communicate with high expectations;
- Respect diverse talents and different ways of learning.

Supported with examples, the above principles are crucial to help students be more actively involved in the learning process than just being passive information receivers.

Especially considering a more conservative nature in our oriental society, the enhanced interactivity and feedback mechanism are important to improve the learning quality and efficiency, to enhance communication, to erase the borders and foster the culture of learning naturally from each other and to better engage the students. Besides the above, with the current development of broadband and wireless networking technology, the capacities of data transmission has been greatly improved, which makes synchronously multi-casting large amount of information possible. It is good time to consider enabling distributed education [11] so that students can choose to learn at a place of their conveniences if they cannot come to the classroom. In the following paragraph, we briefly describe some related studies using Tablet PCs and pen-based technology in education. In an early study [29], Anderson *et al.* show that digital ink may be used to effectively create annotations on the lecture content as a substitute for physical gestures to highlight context and meaning during a lecture. In [5], Anderson *et al.* further developed a Classroom Presenter (CP) tool to better use electronic inks to achieve multiple pedagogical goals in classroom teaching. This presenter tool has been made free and its source is shared on Internet for third-party developments. Recently, the tool has attracted many attentions and more faculties from different universities have start using it to interactively deliver their lectures. We will describe the features of this tool in more details later in this report. In [35], the students and lectures use Tablet PCs together with Classroom Presenter to deliver lecture content and distribute/collect student exercises. While the learning outcomes were not directly evaluated, the student participants showed a high degree to receptivity to the technology. Additionally, it was noted that the inclusion of prepared materials that involved in-class exercise lead to an increase in the preparation time for the lecturer. The work in [10, 18] deploy Tablet PCs in a large-scale engineering educational environment in Virginia Tech.. It should be noted that Virginia Tech. is the first university to incorporate Tablet PCs in large-scale engineering education where each fresh student enrolled in its Engineering school is required to purchase a Tablet PC. The study in [10] addresses a number of related issues in successfully facilitating the infrastructural change including the Tablet PC deployment, pedagogical goals, hardware specification, software tools, pen-based teaching tools, faculty support and training, and system evaluation. Through using both Classroom Presenter and DyKnow vision for interactive classroom teaching, fostering electronic note taking, organization and collaboration using Microsoft OneNote, encouraging electronic homework submission, the new system received positive feedbacks from both the professors and the students. The data analysis shows only after half a semester, all students use their Tablet PCs on daily basis, which include 61% of students who do not have access to computers in their high-school classes. The system also has significant changed the students' learning habits, e.g. note taking and studying. Another study [26] combining Tablet PCs and Classroom Presenter divided a class of 53 students such that 19 uses Tablet PCs, 20 used laptops and 13 do not use any either device. Regarding the overall performance, the study found that the students who utilized Tablet PCs performed slightly better when their final grades are considered. The work in [36] highlighted the use of DyKnow Vision and pen-based Wacom Tablets for the instructor and students in a computer science department for supporting collaborative note taking, classroom interaction, the ability of students to replay lecture material outside the classes, and the basic lecture presentation. The results show the constructed system was helpful in creating an effective learning environment. Instructors reported a heavy usage of the digital ink input during lecture with an approximate 50/50 split in the use the pen and the keyboard input outside a lecture for class preparation. In addition, the study reported that students tended to use the pen and the keyboard inputs in approximate equal parts. Another similar study in [37] investigated the use of digital ink versus keyboard input. In this study, groups of students were given access to laptops or Tablet PCs while the instructor would conduct class lectures using Classroom Presenter. The results of this study showed that Tablet PC users nearly always used the electronic inking option for working on class exercises and performing other simple input operations. However, while students tended to prefer the digital ink option for input in problem-oriented sessions, many students was able to function with either input mode depending on what was available. The work in [38] studied the initiative that would require students to own Tablet PCs. The study utilized a combination of DyKnow Vision and Tablet PCs for the instructors and students. Survey results showed that 90% of the participated students felt the most effective use of the software feature was while using the digital ink input. Students identified their favorite feature to be the ability to see the instructor's notes exactly on their own Tablet and being able to annotate the notes as they wished. Additionally, the survey responses suggested that the students felt there was a positive impact on their learning while using the technology. In most of the above studies, in general, positive survey feedback and perceptions are reported for Tablet PCs' usage together with suitable teaching tools. However, there are also numerous lessons learned from the initial experiences shared in some studies. The work in [18] reports the wireless (802.11) connection issues in a classroom setting when a large number of students, say 250,  attempt to connect simultaneously to the lecturer’s broadcasting stream in a very concentrated lecture hall. The bandwidth demand can be huge. The degraded network service and high latency can quickly lead to student frustration and ultimately the abandonment of connecting to the instructor’s application. Special care needs to be taken in designing wireless network in the classrooms to enhance the wireless transmission capacity. The work in [8] investigate the first use of Tablet PCs to Engineering design course to a group of students, who are mostly new to Tablet PCs. From the survey, though the use of Tablet PCs find some support to certain extent, the study suggests that there should be activities or exercises targeted for students to make full use of the unique features Tablet PCs. And it is better to perform necessary training to equip the students with necessary knowledge and skills to use the Tablet PCs’ inking features and the relevant tools more effectively. The work in [19] studies how students with different learning styles benefit from the first-time incorporation of Tablet PCs in learning a math and an electromagnetic theory courses. The survey results showed that the visual/kinetic learners (those who are both good at learning through seeing and doing) benefit most from the new course design and the kinetic learners (those learn best through doing) perceived least benefited from the changes. Several lessons learned include: 1) The Tablet PCs are felt a bit restrictive in some cases and both students and teachers prefer mixing Tablet PC exercises together with Blackboard works; 2) For note taking, the skeleton notes shown by the lecturer should be made very brief to only outline the learning material. This gives room for students to enrich it; 3) It could take a great deal of time and effort to convert the existing lecture notes and convert them for the use on Tablet PCs. Though it is generally agreed that Tablet PC is an effective learning tool, properly preparing the material by professor is a crucial step for the effectiveness of using Tablet PC technology. A recent study in [13] summarizes their initial experiences in promoting Tablet PCs to support engineering study. Some highlighted issues include: 1) the stylus pen should be properly calibrated to show its location on the screen; 2) To capitalize Tablet PC inking features in some classroom activities, it is important to ensure everybody has a tablet; 3) both students and instructors need to have some training; 4) Mobility of the lecturer can be constrained due to a physical cable connection; 5) It is good to keep the blackboard; 6) It is helpful of being able to make free-style annotations. Deploying proper software to have such a feature is beneficial; 7) Computing devices can be a distracting element to students during the time that they are not used in a lecture.

Besides the above studies, a number of new teaching software applications have been developed to facilitate the specific pedagogical goals in classroom settings. These tools include the classroom presentation management tools such as Classroom Presenter [5, 30], DyKnow Vision [31] and Ubiquitous presenter [24]. All these tools support electronic inks, real-time ink broadcasting, in-class interactive activities and anonymous work submission. The main difference is that Classroom Presenter and Ubiquitous presenter are free tools while DyKnown Vision is a commercial tool. Classroom Presenter is a window-based point-to-point type of application through IP, which has all the basic functionality to deliver electronic inking-friendly interactive lectures. Currently, this tool is maintained and further developed by University of Washington. Developed by University of California, San Diego (UCSD) from the nutshell of Classroom Presenter, Ubiquitous presenter inherits most of the functionality of Classroom Presenter except that it is a web-based application. It runs on heterogeneous machines, which support JAVA platform, including many mobile devices. Only the lecturer needs to install the software in order to run an online lecturer. However, to run a Ubiquitous Presenter application, one still need to use the server provided by UCSD, which could be inconvenient in our case, as our courses are taught in Singapore. As mentioned earlier, DyKnown Vision is commercial tool, which has nearly all the features that Classroom Presenter has. Additionally, it has voice recording and ink-replay features which allows students to review the archived lecture. DyKnown uses a point-server-point based connection architecture. The powerful server in the middle tier greatly enhances the processing and data exchange capability for a very large-scale lecture, e.g. for teaching a thousand of students. While at the same time, Classroom Presenter supports up to about 80 students in a lecture in a recent test conducted in Virginia Tech. Besides these classroom presenting tools, several other reported tools developed include WriteOn [17, 32], OrganicPad [25], BIRD note-taking system [34] and MessageGrid [39]. The WriteOn tool is developed to help explain the materials during a lecture through ink annotations. By adding a new transparent writing layer onto the desktop screen, an instructor can write anywhere to explain a concept e.g. explaining the functions of different coding blocks in software engineering course. The captured inks are multi-casted to students in a real time and the instructors can also video record of the inking process. OrganicPad is a sketching and recognition tool to make drawing and teaching of organic chemistry easier. The BIRD note-taking system allows a faculty to annotate easily during a student's presentation. And the MessageGrid incorporates artificial intelligence to help instructors better identify the misconception by automatically analyzing the replayed ink sequence. Besides the above new tools, which are often targeted on very specific disciplines, some commonly used general purpose tools have also incorporated the electronic inking features, which can be used to enhance the learning environment. Bundled with any Microsoft Tablet PC OSs, Windows Journal software is an electronic notebook which allows free annotation, sketching and drawing. Still under development, the Microsoft education pack contains a collection of learning tools including equation writer, Tablet PC ink flashcard and so on. Since Microsoft Officer Version 2003, electronic inking features have been built into Microsoft Word, Excel and PowerPoint for easy annotating in these documents or during a PowerPoint presentation. Microsoft OneNote is an advanced note-taking tool, which enable ink recognition, intelligent organization, student collaboration and ink-based searching. Adobe Acrobat also added inking and tagging features to allow ink annotations in PDF documents. Nowadays, the teaching activities can easily require real-time exchange of large amount of multimedia data streams including video and audio data. To optimize the network efficiency for multimedia streaming, Microsoft research has also developed shared-source platform, named ConferenceXP [40], to provide simple, flexible, and extensible conferencing and collaboration using high-bandwidth networks and the advanced multimedia capabilities of Microsoft Windows. Such platform together with Classroom Presenter has been tested to conduct interactive distributed lecture in 4 different venues, where videos, audios and electronic inks from different venues are exchanged.

The previously reviewed numerous studies have shown various positive impacts of introducing the Tablet PC technology into higher education. Compared with the usage of normal notebook PCs, the usage of Tablet PCs are still in its infancy and its innovative uses can still be explored in education of different courses. Though from the lessons learned, we understand that technology imperfections such as wireless networking capacity, restrictions on free writing and artifacts of the software tools can negatively affect users' experience and ultimately affect the adoption of the new system, such problems can be improved in the future tablet PC technology and precaution can be taken in the implementation to bring down their negative impacts to its minimum. Currently, there are already some Tablet PC educational tools available and we expect that in the future, these tools will be improved based on the current lessons learned and the new tools for discipline-specific education will emerge. This provides new opportunities to continuously development and improvement of our learning framework. After studying the experiences in other universities, we have set up our pilot Tablet PC-based teaching environment with an aim to enhance the pedagogical goals of active, student-centric and interactive learning and minimize the possible negative user experiences.

This chapter is organized as follows. In the section 2, we describe our teaching environment in details. Section 3 describes some limitations and some possible future extension. Section 4 gives the conclusion.

## 2. Tablet-PC Based Learning Environment

Prior to this development, notebooks PCs have already been widely by faculties in our school of electrical and electronic engineering (EEE) in Nanyang Technological University (NTU) for lecture or tutorial presentation. Microsoft PowerPoint is the most frequently used presentation tool. The high-speed broadband network and WiFi are available for Internet access. The faculty staffs have been given choices to use either Tablet PCs or high-speed desktop PCs for their daily works. Hence, a portion of faculties had prior experiences of using Tablet PCs. Though many student had not used Tablet PC before, they generally possess good computer-related knowledge since many subjects, design labs or project assignments require students to work on PCs and search for information on Internet. The electronic learning system, named edveNTUre, is commonly used for instructor and students to share and exchange course-related information and materials. To encourage new ways of learning, NTU has also supported faculties to change their teaching style with new technologies, e.g. using CLICKER devices in a lecture, which allows students to give instantaneous feedback on multiple choice questions. In this environment, the EEE school management has also supported our project by identifying the ICIS Practicum lab for our construction the Tablet PC based learning lab.

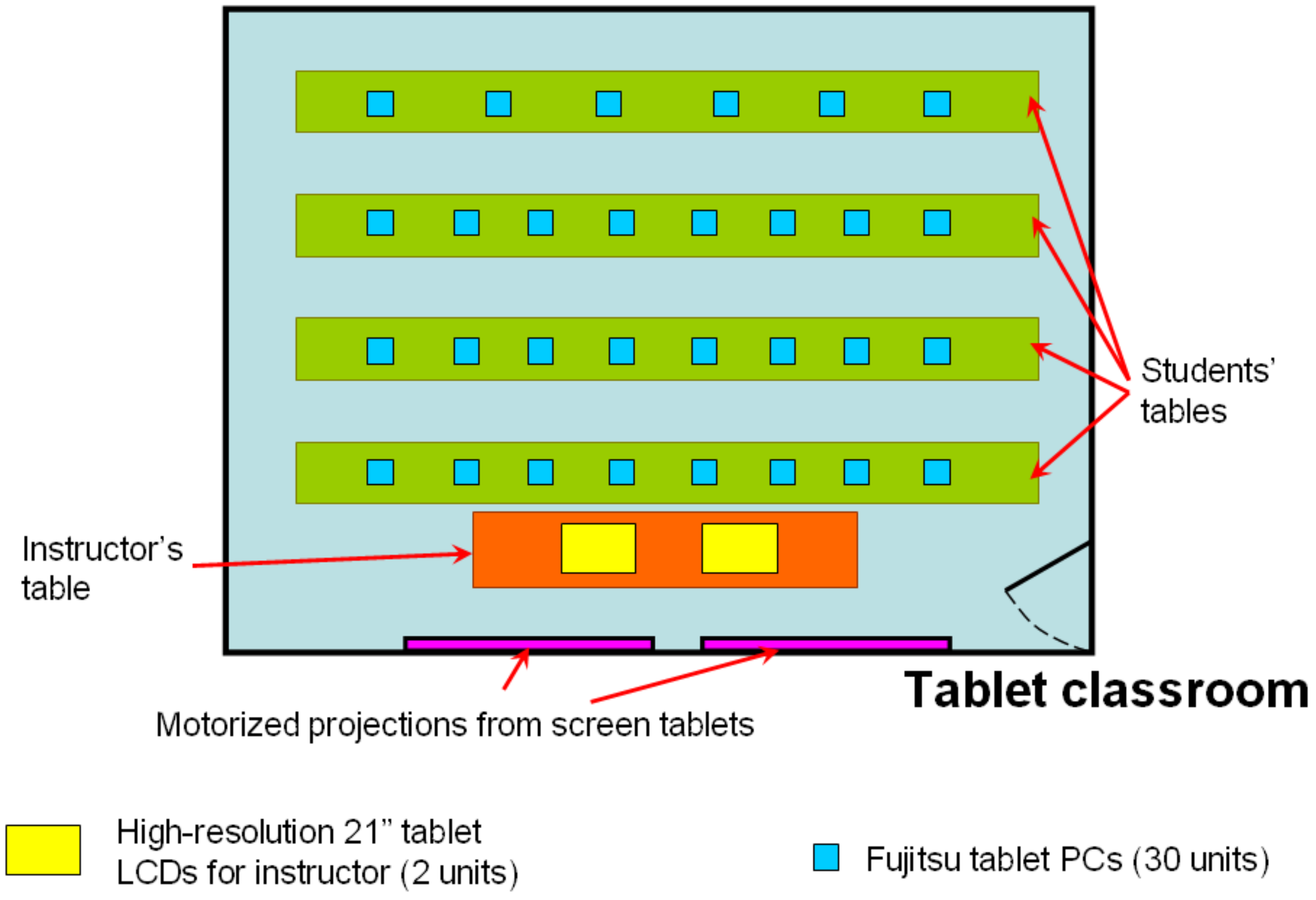

Fig. 2 Top-View of Our Established Tablet-PC Based Learning Environment

By reviewing the literature of similar development in the previous section, we understand that the proper deployment of the new technology has been the key factor to its final success. Hence, we redesigned the ICIS Practicum lab to incorporate the Tablet PC technology to mainly support the following goals:

Goal 1. Easier ink annotations with least constraints;

Goal 2. Enhanced active learning with timely feedback;

Goal 3. Enhanced student collaborations.

Goal 4. Lecture recording

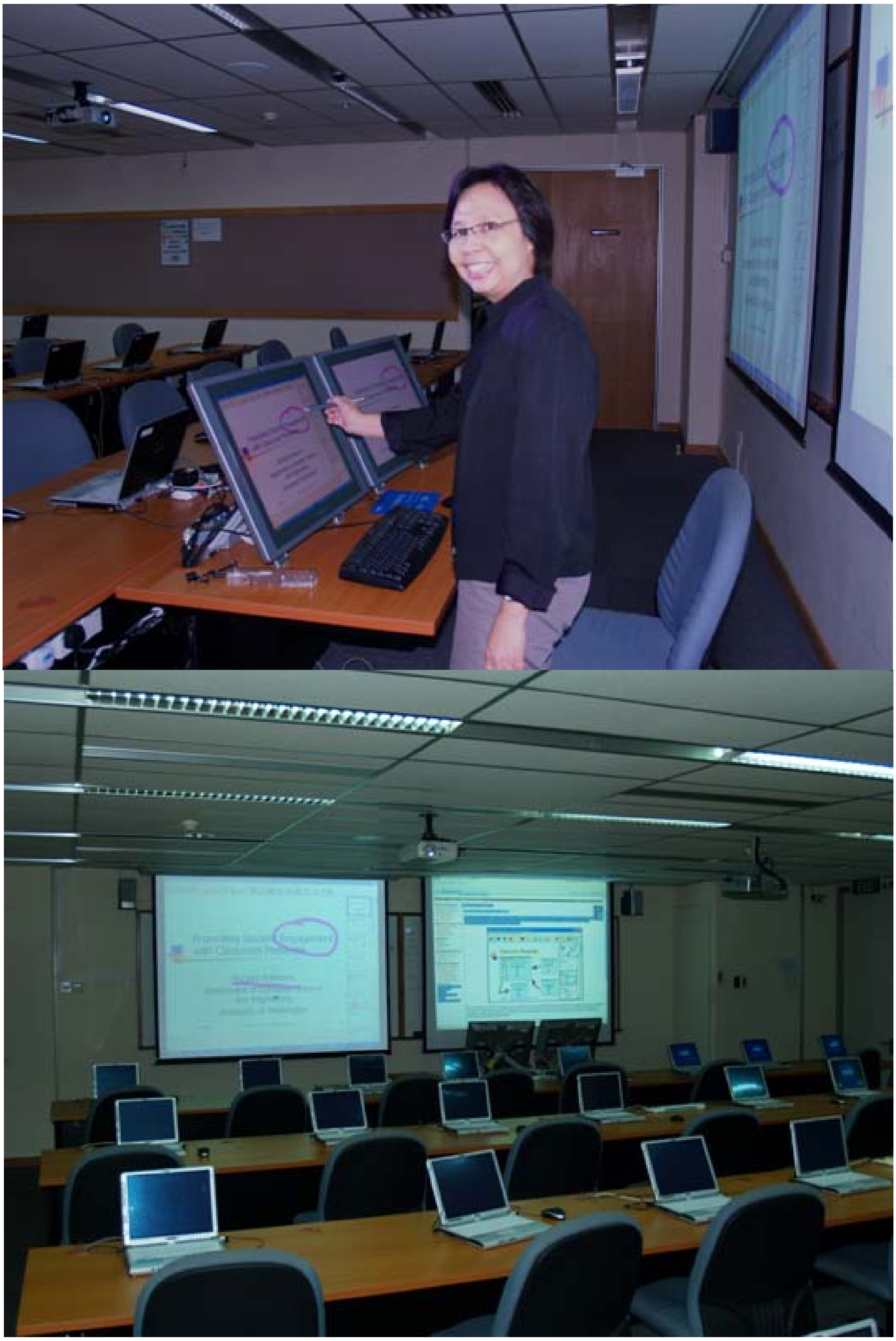

Fig. 3 Photos of Our Tablet PC Based Learning Environment

Our classroom is designed to support a class size of 30, which is sufficient for normal usage as a small lecture room and online laboratory. The top design view of our established Tablet-PC environment is shown in Fig. 2 and two classroom photos are shown in Fig. 3. The classroom is equipped with 30 Fujitsu Tablet PCs, each for one student, 2 projectors and motorized screens, 2 Wacom Cintiq 21-inch screen tablets and the blackboard is maintained behind the 2 motorized screens. Shown in Table 1 and Table 2 are the hardware configuration and the software tools equipped for the Tablet PCs and the lecturer's desktop PC respectively. Below we discuss how the classroom supports the pedagogical goals.

Table 1 Tablet PC requirements

| Item | Details |
| --- | --- |
| Brand and Model | Fujitsu Lifebook T4220N2 |
| Platform | Tablet PC convertible |
| OS | Windows XP Pro Tablet PC Edition |
| Processor | Pentium Core 2 Due 2.2 GHz, Chipset T7500 |
| Display | 12.1-inch XGA Transmissive TFT |
| Memory (RAM) | 2 GBytes |
| Hard disk | 120 GBytes; SATA HDD |
| Video card | 128 Mbytes |
| Optical drive | Dual layer DVD/CD Multi-Writer |
| Input/Output | USB 2.0 |
| Wireless | 801.11, Intel Pro |
| NIC/Ethernet | 10/100/1000 Ethernet |
| Warranty | 3 years onsite warranty |
| Software tools | Microsoft Office 2003, Microsoft OneNote, Microsoft ink desktop, Adobe acrobat, Microsoft education pack, Microsoft software experience pack, collaborative office tools, classroom presenter, … |

Table 2 OS, Special Hardware and Teaching Software Tools for the Lecturer's PC

| Item | Software Tools |
| --- | --- |
| Brand | NEC Powermate |
| Platform | Desktop PC |
| OS | Windows XP Pro Tablet PC Edition |
| Display Card | NVIDIA GeForce 8500GT 512 MB with Dual-Monitor support |
| Monitor | 2 Wacom-Cintiq 2100 AG0C Tablet Monitor |
| Tablet Driver | Wacom Tablet Driver v6.08 |
| Teaching tools | Microsoft Office 2003, Microsoft ink desktop, Adobe acrobat, Microsoft education pack, Microsoft experience pack, classroom presenter, PPT plugin for classroom presenter, collaborative office tools and Justwrite Office, Camtasia Studio… |

## Goal 1-Inking Anywhere

To support the Goal 1, we need maximize the electronic inking capability and make electronic inking comfortable with very limited constraints. This is achieved through both hardware configuration and deployment of selected set of electronic inking tools.

As a central role of classroom teaching, it is particularly important to allow the instructor to write freely onto the screen. We equip the lecturer's PC with two Wacom 21-inch Tablet monitors, which are high-end products to provide large tablet area, high data rate and high density of transducer sensors. These qualities, in reality, give the instructor not only large writing area but also excellent "writing feel" as if he/she is writing on a real paper. Moreover, tablet screen can be easily adjusted in case the instructor needs to change a writing posture. The two Tablet monitors are both connected to the instructor's PC. By configuring dual-monitor setup and properly calibrating the pen locations, the instructor can use the monitors as a large combined screen. The 2 tablet screens are separately connected to the two projectors to help the instructor simultaneously display more information. By default, the Tablet screens are separately projected onto the two motorized drop-down screens, however, the instructor can easily turn off one or both projection and wind up the motorized projection in the case he/she needs to use the blackboard behind using provided marker pens. It should be noted that in the old days that many lecturers liked to write on the multiple blackboards in turns so that students can still see the previously finished blackboard while the lecturer is writing on a new blackboard. Similar electronic configuration in the proposed system can be easily set up on our dual-projection screens so that students can still see the previous slide while the lecturer moves on explain a new slide.

The Windows XP Tablet PC edition provides the OS-level support of electronic inking features and most of the teaching software tools equipped as shown in Table 2 provide good electronic inking features. It should be noted that there have been several latest Tablet PC OSs available including Windows 7 and Windows Vista. However, since existing engineering courses requires installation of some other teaching tools which have compatibility issues with these new Tablet OSs, we choose to install the earliest Windows XP Tablet PC OS. Below, we highlight several typical usages of electronic inking in a lecture.

## Annotation in a Presentation

As illustrated in Fig. 4, the Classroom Presenter (CP) tool has a pen-centric interface, which allows the lecturer to annotate freely, to navigate slides easily and to preview slides during a presentation. At the beginning of the class, the lecturer simply starts his lecture by loading a PowerPoint presentation slides. These PowerPoint slides can be prepared beforehand using the Microsoft Powerpoint. Students can connect to the lecturer's server application through IP using their Classroom Presenter client application. While the lecturer annotate onto his slides, the electronic inks are multi-casted to the students' client application at real-time.  At the same time, the student can still take their own electronic-ink notes onto the slides and choose to save both instructor's inks and their own inks together with slides at the end of the lecture.

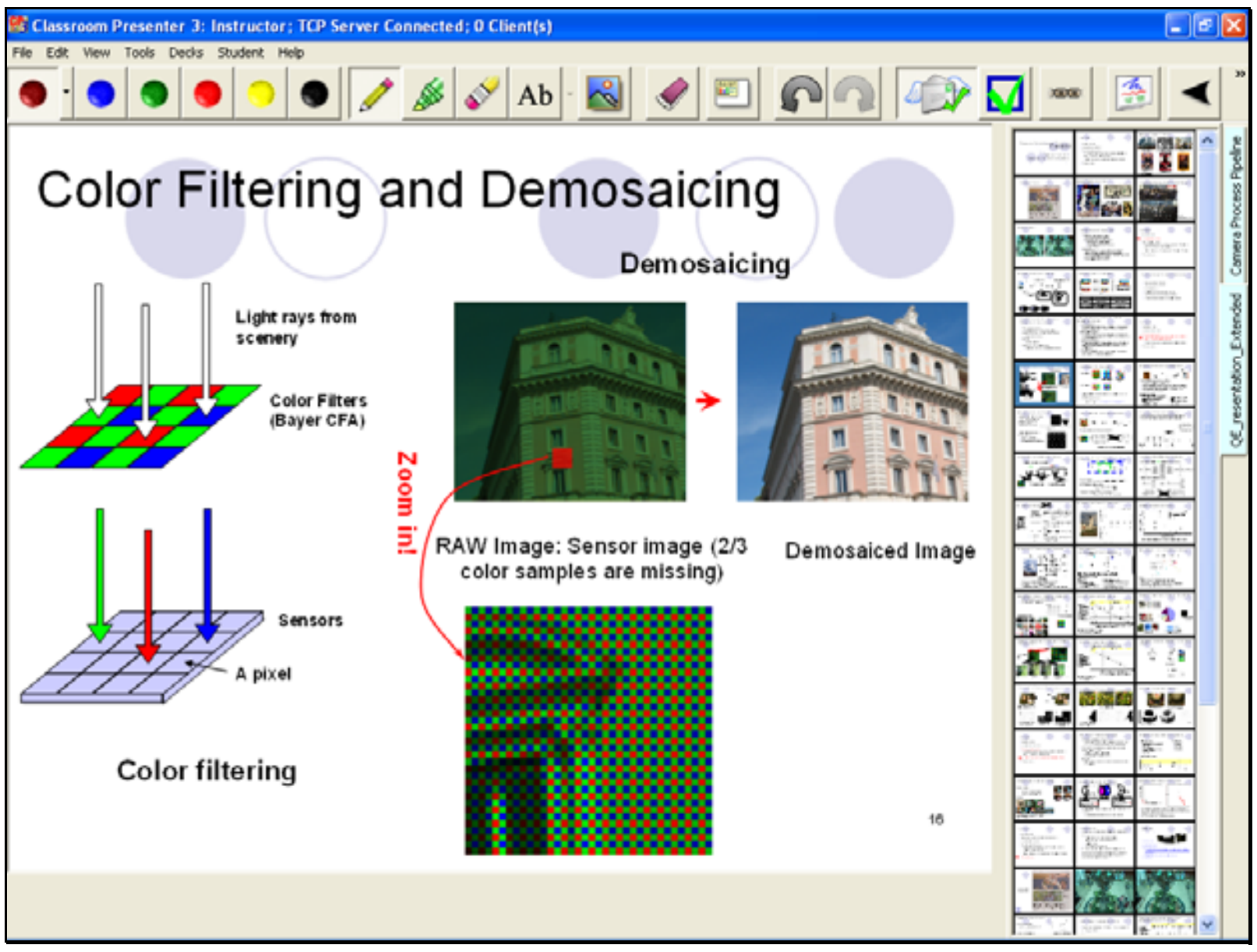

Fig. 4 Pen-Centric Interface of Classroom Presenter

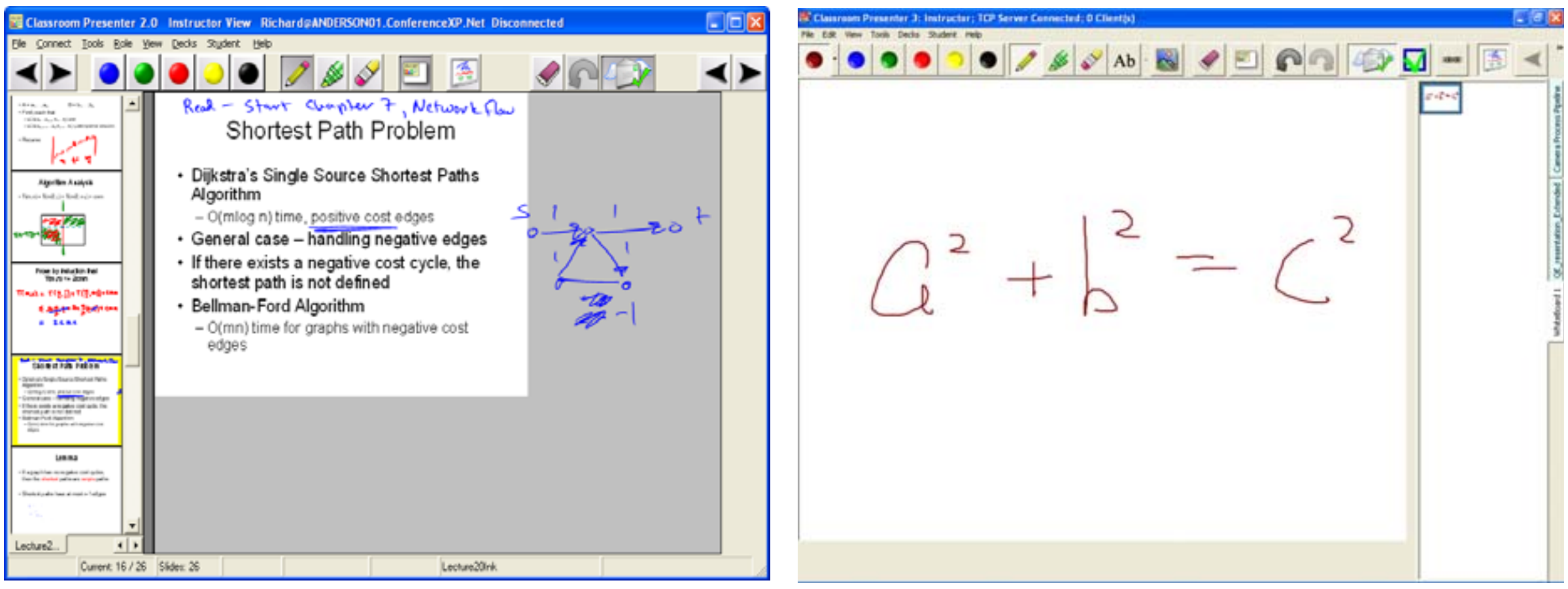

Fig. 4 Make Spaces for Pen Annotation by (a) Shrinking the Slide area and by (b) Inserting Blank Slides

As shown in Fig 5 (a), the CP tool also allows shrinking the slide area to certain extent to make more empty space for electronic inking. Further, the lecturer can choose to insert blank slides for more space of inking.

These above features will make the electronic inking a comfortable task. The pen-centric interface also makes the lecturer feel easy to complete the common tasks such as slides navigation, previewing, menu selection, change of pen color, width and types, and erasing, etc. using a pen. This minimizes the frequency that the lecturer needs to switch to other input modalities such as keyboard and mouse.

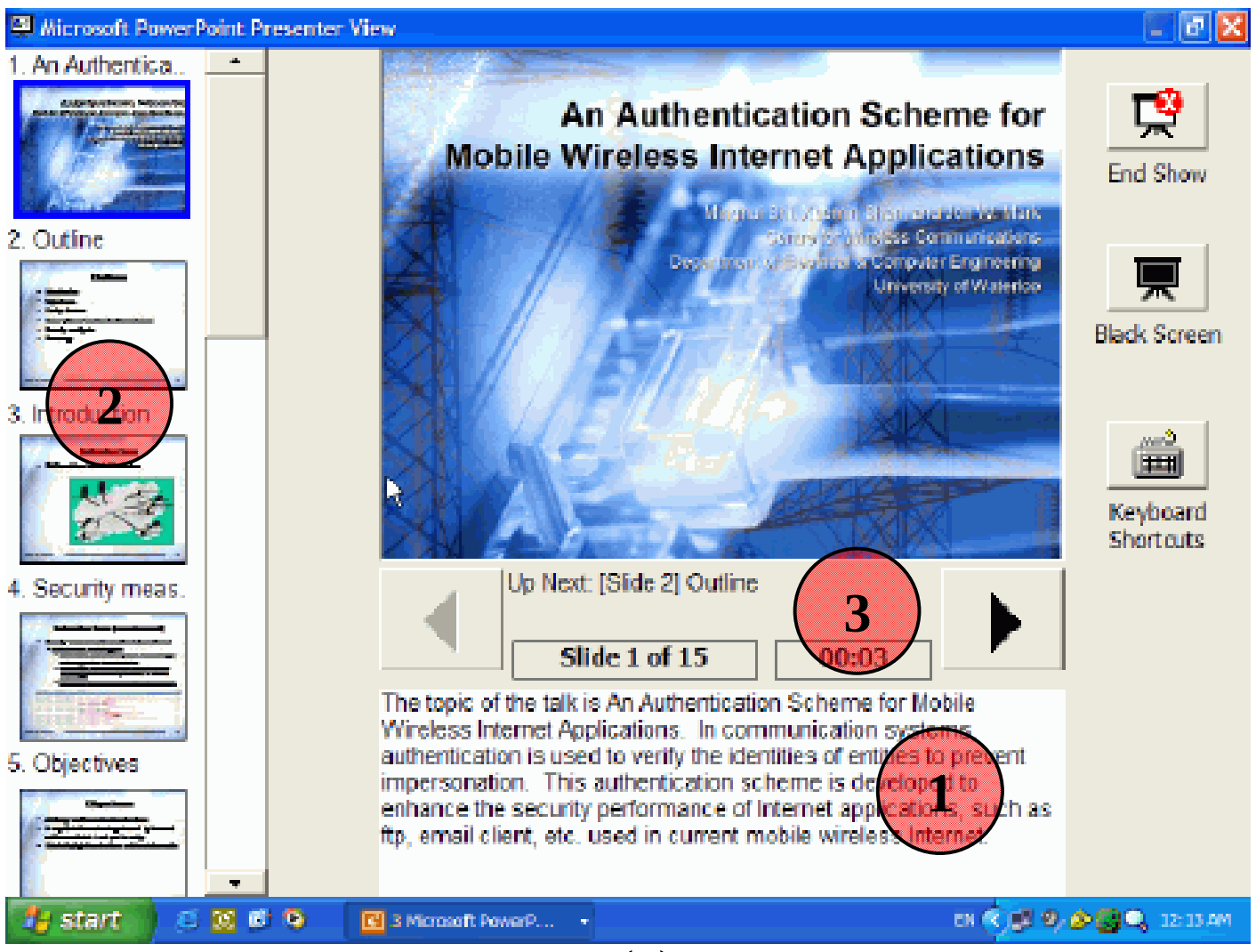

(a)

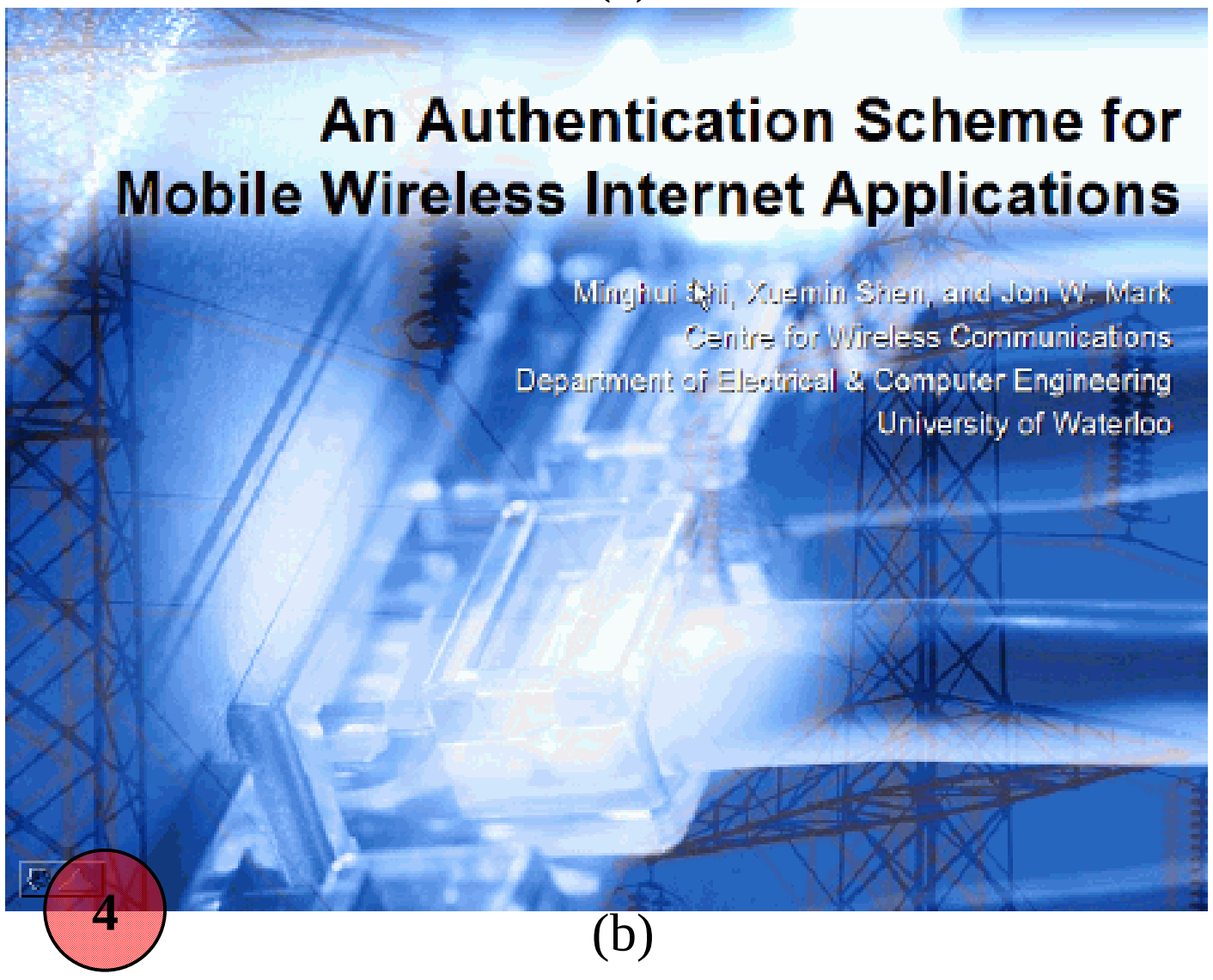

(b)

Fig. 6 Presenter's View in (a) and the Projected View in (b). The circle with a number in it indicates: (1) Slide notes; (2) Slides thumbnail; (3) Duration of a presentation; (4) Inking pen drop-down menu

In the case that the lecturer prefers to using Microsoft PowerPoint to conduct a lecture, we recommend the lecturer to use presenter's view to enhance the lecture. As shown in Fig. 6, in such a scenario, the lecturer will see a screen with much richer information than the projected view that students see. In the presenter's view, the additional information such as the slide notes, the slide thumbnail, the presentation duration and oversized control button helps the lecturer with better delivery and time control. It should be noted that the PowerPoint presentation also support free e-ink annotation. The lecturer can select different pens in a dropdown menu located at the bottom-left of the projection screen for annotation or highlighting purposes. Additionally, the lecturer can choose to use JustShow tool in Wacom JustWrite office to deliver the e-ink based lecturer. The JustShow has a pen-writing friendly drop-down toolbar to allow rich pen-based activities including annotation, drawing and sketching. Comparatively, it has more inking-related features than Microsoft PowerPoint 2003, but the limitation is that it cannot be used together with the Presenter's view.

## Screen Annotation

Besides presenting PowerPoint slides, we also note that sometimes it is important to be able to writing directly on top of the desktop screen over any running applications to explain some concepts. For example, when a lecturer needs to explain the different functional blocks of some source codes contained in a running Visual Studio application, it is often good if the lecture can highlight some source codes and write some annotations nearby using electronic pen. It should be noted in most of software application, e.g. Visual studio, electronic inking feature has not been built in. To address the inking problem in such a scenario, we recommend lecturers to use the Screen Markup tool in the Wacom JustWrite Office for such cases. The screen Markup as shown in Fig. 7 provides the transparent writing layer so that one can write on top of the existing applications. Though Microsoft ink desktop as shown in Fig. 8 and WriteOn tool can do similar things, we prefer the Screen Markup tool because it works better in the case when multiple monitor screens are combined.

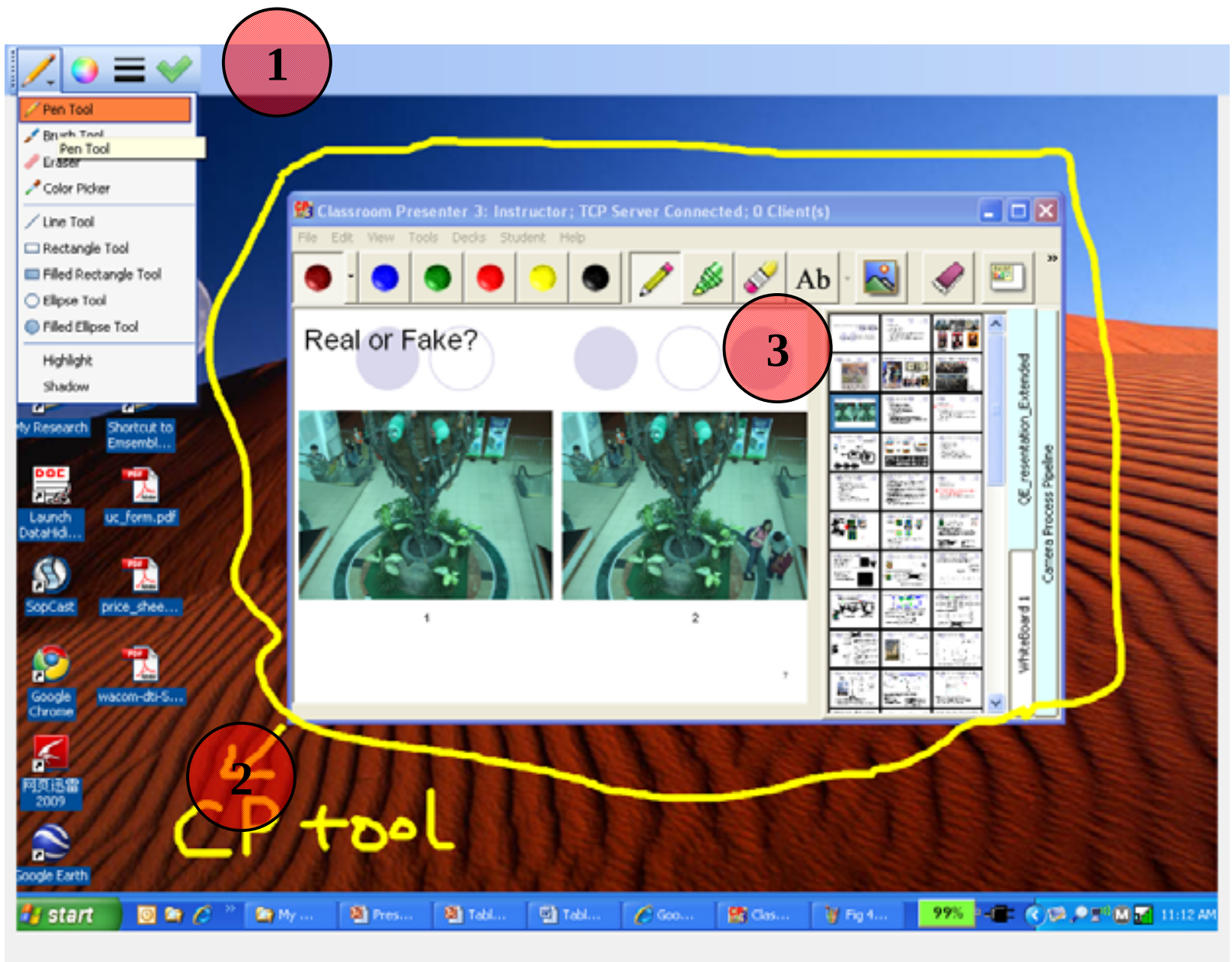

Fig. 7 Annotating on the Screen with the Wacom Screen Markup Tool. The circle with a number in it indicates: (1) Screen markup pen control toolbar; (2) screen markup ink; (3) background application

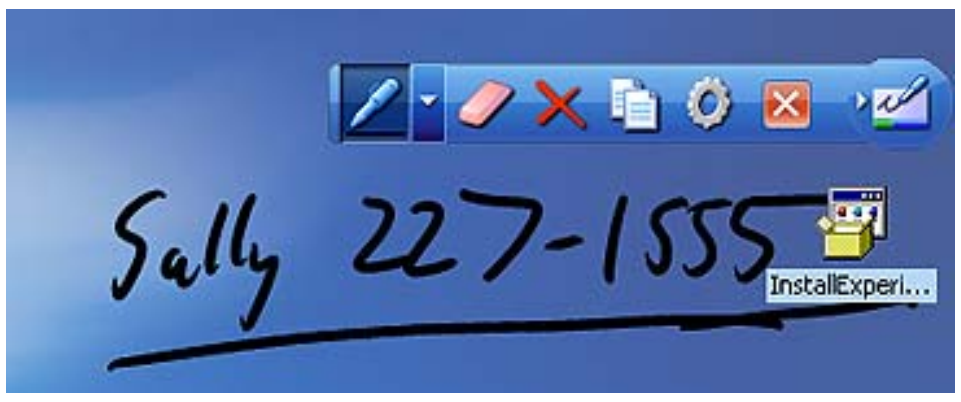

Fig. 8 Annotating on the Screen using Microsoft Ink Desktop (Source: Internet)

## Inking in Documents

Currently, nearly every undergraduate student in our Engineering school prepare their assignment or report in electronic format, e.g. as office documents or PDF documents. In the future, we expect that all electronic reports are also submitted electronically, which will save a lot of papers and printing cost. The lecturer grades these reports electronically and returns marked report back to the students. Hence, it is important that our tools support electronic inking in the documents.

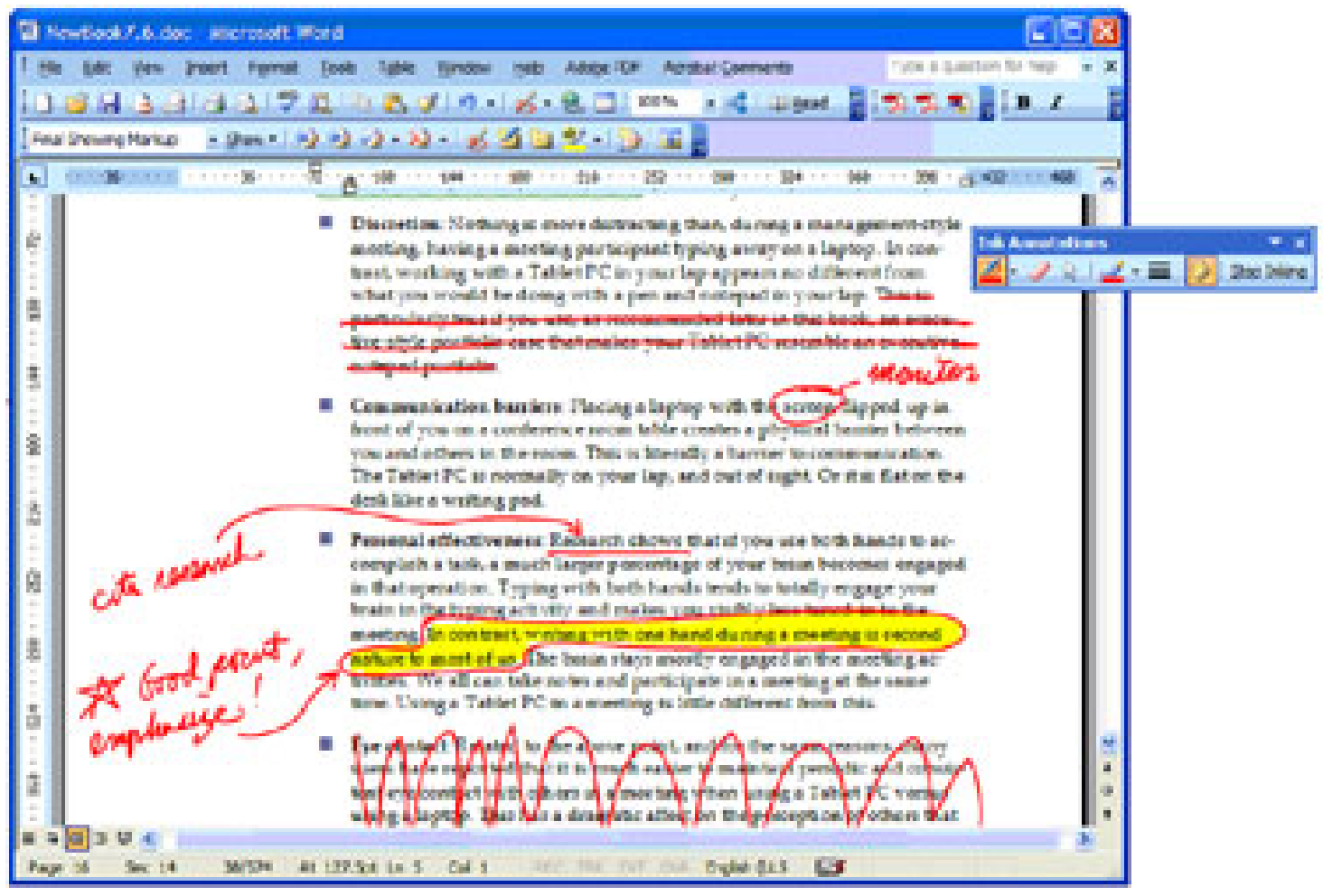

Fig. 9 Annotation in a Microsoft Word Document (Source: Internet)

With the electronic inking support at the OS level, since Microsoft Office 2003, electronic inking in Word is supported. Fig. 9 shows an example of ink-annotated Word document. Similarly, the latest Adobe Acrobat has the Pencil tools which allow basic annotations in a PDF document. Since most of electronic documents are either in Microsoft Office formats or stored in PDF format, availability of these tools will make electronically grading different types of student’s reports and assignment an easy task.

In some occasions, it is desirable to be able to secure the electronic inks together with document container to avoid possible disputes since the ink data can be tampered easily with no traces left. In our recent work [41, 44], we have proposed a solution to use lossless data embedding technique together with public-key infrastructure cryptographic techniques to secure electronic inks. The integrity of the electronic inks can be protected within the document container and at the same time, the original ink data can be losslessly recovered.

## Goal 2-Active Learning with Timely Feedback

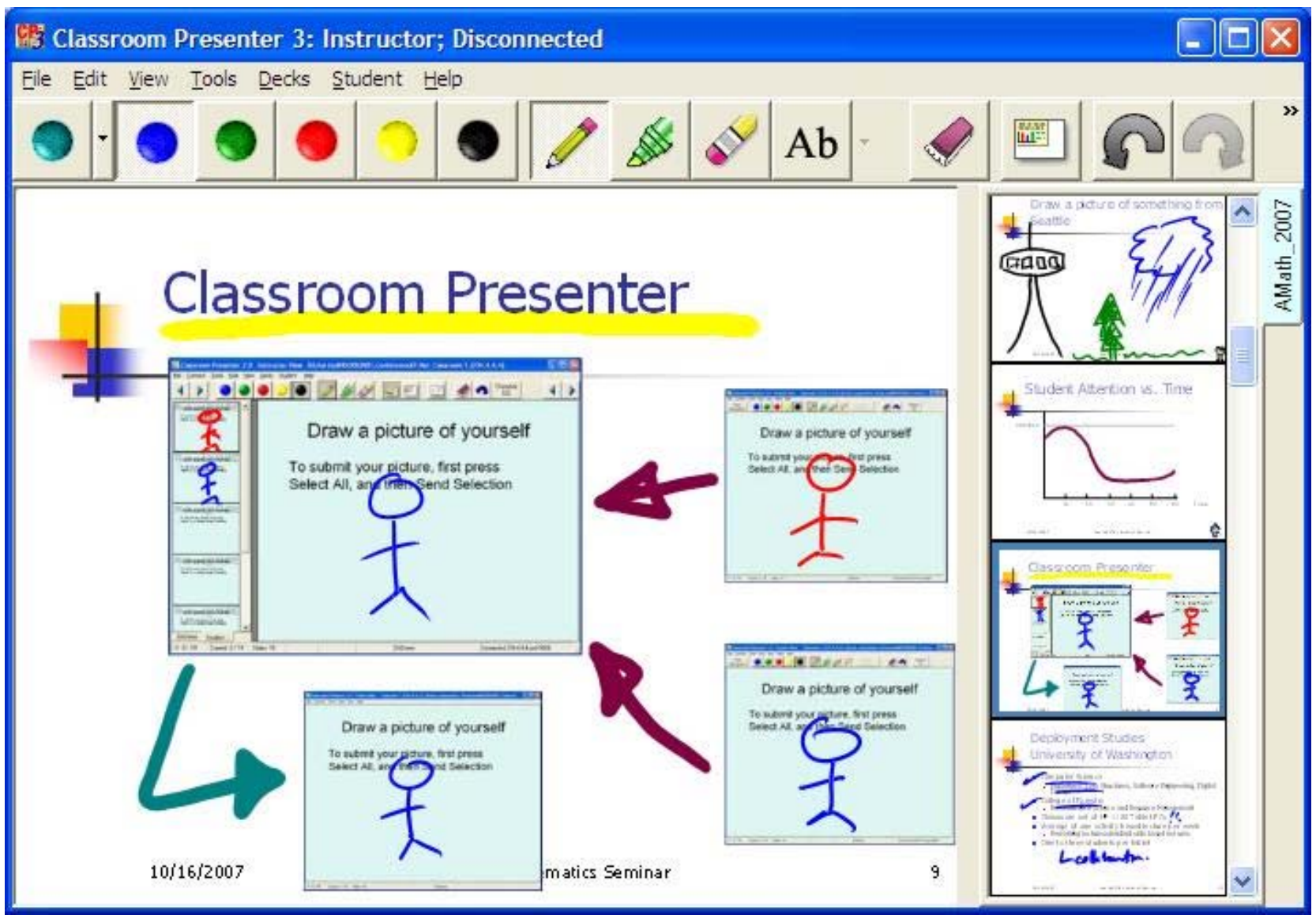

Fig. 10 Illustration of Student Submission using Classroom Presenter

The core elements of active learning are student activity and engagement in the learning process. This is often contrasted with traditional lecture, where the lecturer does the one-man show and students act as the passive information receiver. To make the students more actively involved in the classroom learning process, the portion of student activities should be increased and this will also strengthen the knowledge presented, timely correct the common misconceptions and improve the student engagement. However, since there are many students but only one instructor in a class, such student activities can be time consuming to implement and it is also very difficult for the lecturer to quickly extract meaningful information from the large amount of student works. The Classroom Presenter tool has provided some remarkable features to address the above difficulties including the student submission and polling features.

As illustrated in Fig. 10, a lecturer can design and perform student activity easily using classroom presenter while teaching in a classroom.  Firstly, the lecturer prepares the activity slides, which contains the problem to be solved or an exercise to be done in the class. The activities can be placed in such a way that in every 15-20 minutes that student would need to do some in-class exercise and submit their work to the teacher. During the teaching, when comes an activity slide, the slide would be automatically sent to the students' client classroom presenter application. The student complete his exercise on the empty space of the received slide using a tablet PC and the stylus pen, then simply press a "submission" button to send his work back to lecturer. Within a given duration when many student works are received, the lecturer can quickly browse through these works, spot the common misconceptions, and publicly show and discuss the selected typical student works in order to strengthen the knowledge and principles explained. It should be noted that the student submission is done in any anonymous manner to mainly protect the privacy of the submitted work. This feature assured that even after publicly displaying the mistaken solutions from some students, they will not be discouraged to submit their work during the next activity.

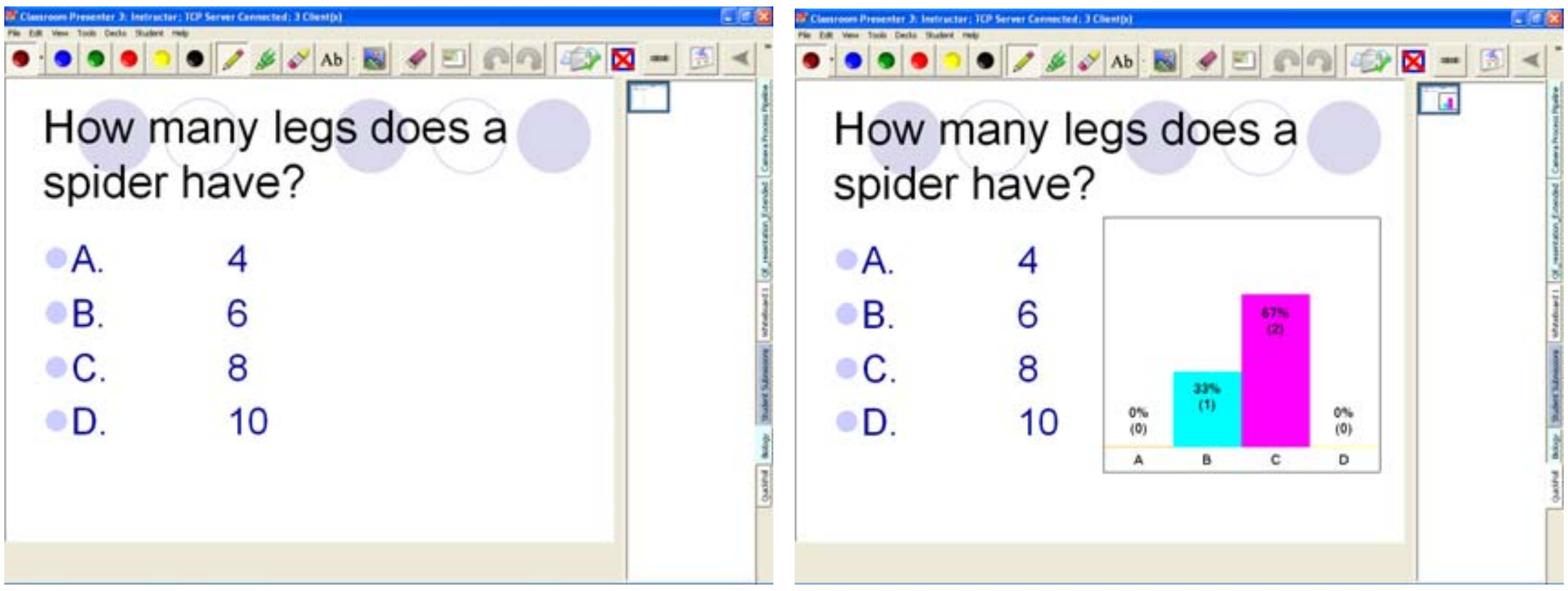

Fig. 11 Polling with Classroom Presenter. On the Left is the Polling Question Slide and on the Right is Distribution of the Result Received.

Sometimes, browsing through the students' submitted work can be laborious work especially if the number of students is large. In this case, the lecturer may consider using design multiple choice questions and use the polling features of the Classroom Presenter to quickly get the statistics of the students' responses. Shown in Fig. 11 is example of polling. With simple ABCD type of multiple choice, the instructor can quickly access to the understanding the students. Besides ABCD types of multiple choice questions, there are many other types of multiple choice questions are available for a lecturer to choose.

As we can see from the above examples, making the lecture interactive is easy with Classroom Presenter. The question slide is automatically transmitted to all students' Tablet PCs through Internet. Students only need to focus on how to best answer the question instead of copying down what is shown by the lecturer first. The submission process is also easy, which requires a simple click on the submission button. This minimizes the waste of time and energy due to the overhead processes and hence makes the classroom learning more concentrated, fruitful and with better control. Also due to the conservative nature in Asian classrooms, many students feel uncomfortable to raise their hands and open their mouths to ask questions or give the lecturer feedbacks directly even if they are asked by a lecturer to do so. To encourage the student feedback, the lecturer can choose to make "allowing student submission" open all the time to the students. A student can anonymously write down their questions onto a slide and send it back to the lecturer instructional application at any time. The lecturer can check the student submission container once a while in the class and upon seeing the student feedback, the lecturer can quickly address it and this would ensure problems arising can be timely addressed to have a good flow of student learning process.

## Goal 3- Learning Collaboratively

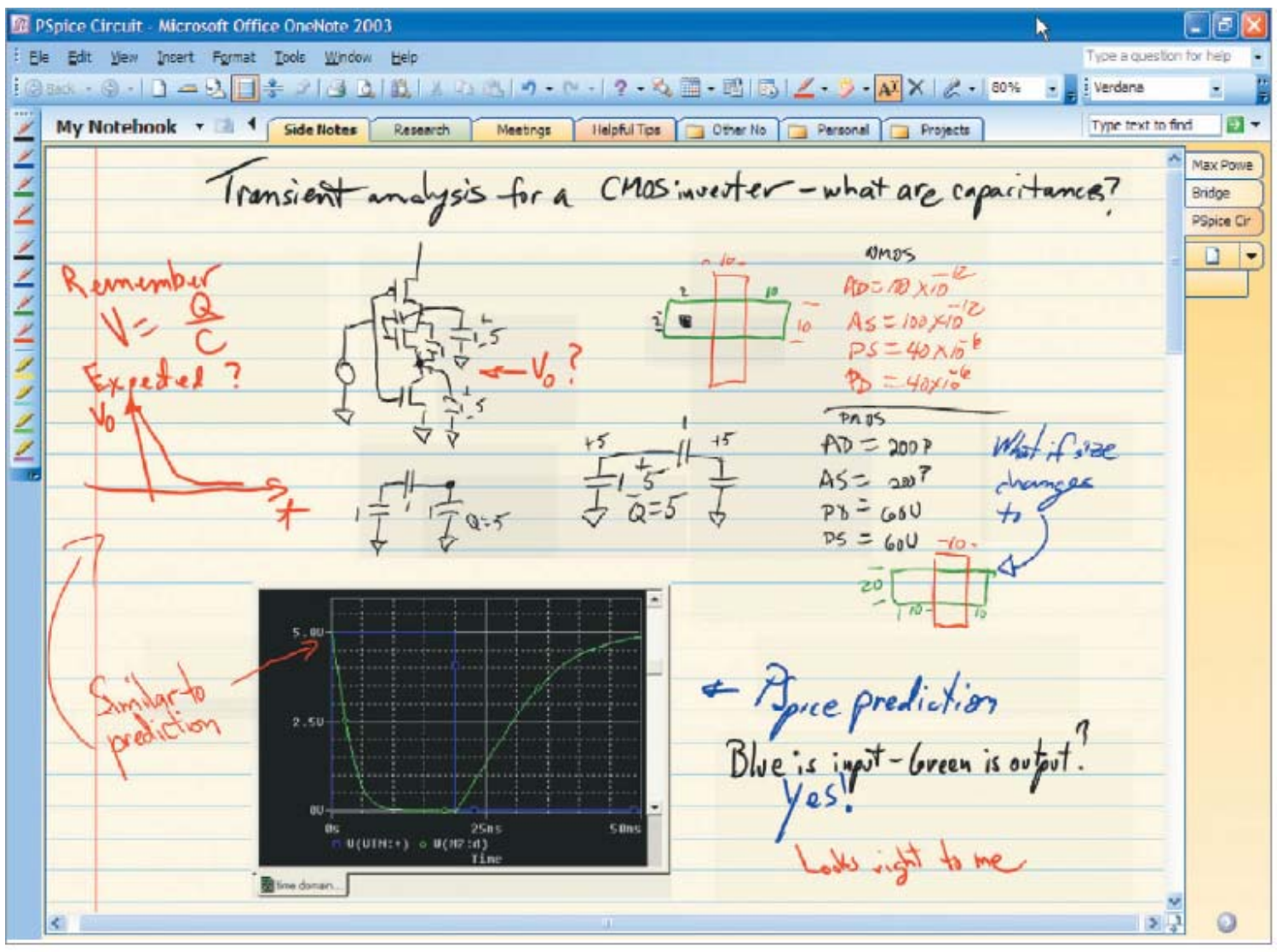

Fig. 12 Using Microsoft OneNote to Collaboratively Solve a Problem [10]

Collaborative learning requires two or more students to work together on a same topic as opposed to the individual work. Past research has shown that collaboration usually improves desirable learning outcomes in academic programs, such as academic achievement, interpersonal skills, self-esteem and retention of the knowledge learned. Our classroom setup also well supports such collaborative learning.

In a group-based exercise, usually it is good that the group members sit closely and work together on the same piece of work. Though currently all our Tablet PCs are tied with network cables and power cords, if needed, such wires can taken out for a long while to run the Tablet PCs on batteries and the wireless connection so that each student can carry his tablet PC around and sit closely in his/her group for a discussion. There are several tools to allow several students to collaboratively working on the same applications. As shown in Fig. 12, with the OneNote tool equipped, students can work on the same application collaboratively. Once a student makes a change, the change is immediately reflected into other students' OneNote application. Besides OneNote, we have also installed the CoOffice tools for collaborative learning as shown in Fig. 13. The toolbox includes CoWord and CoPowerPoint so that multiple students can collaboratively work on the same Word or PowerPoint documents together at real-time.

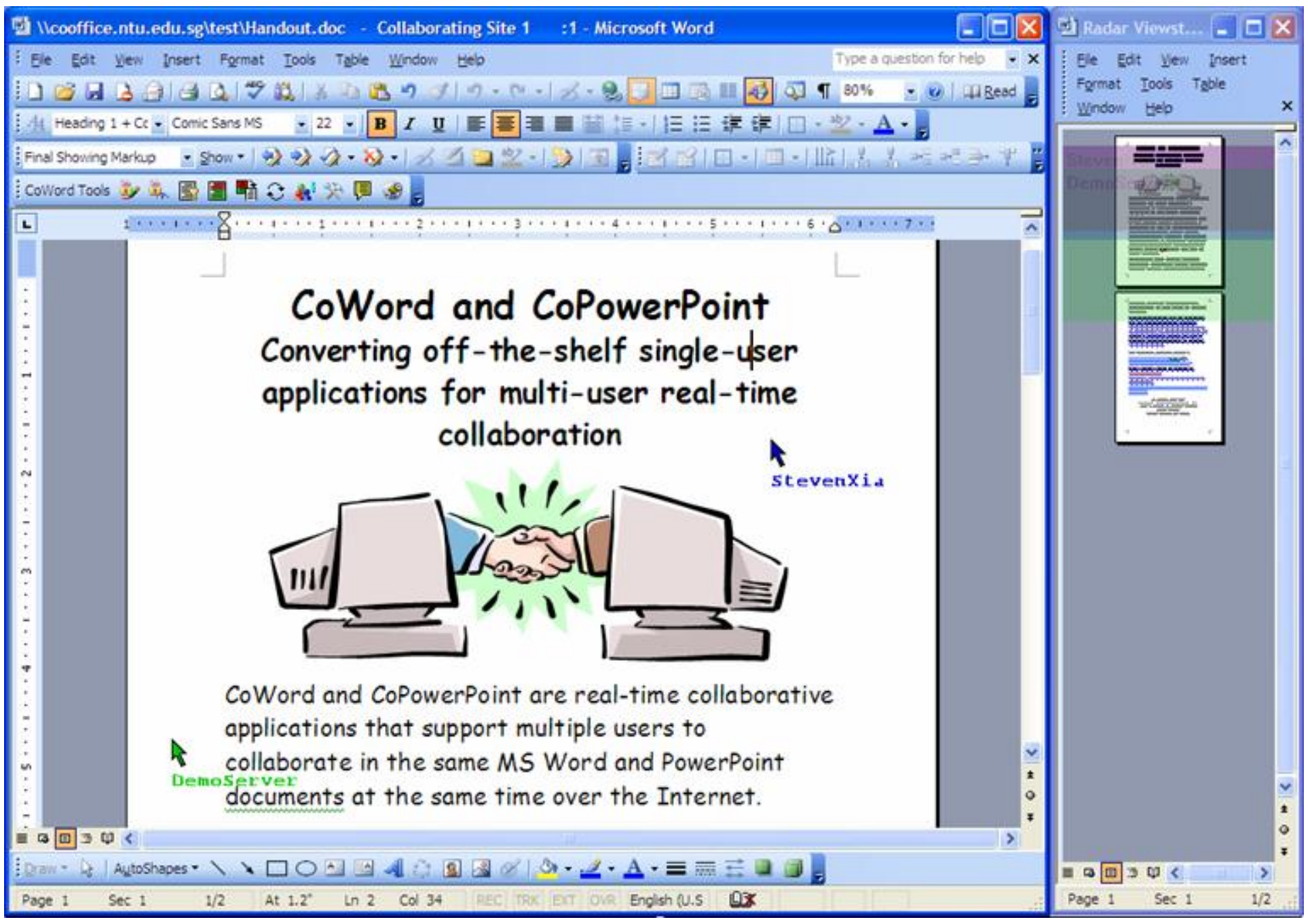

Fig. 13 CoOffice Tools for Real-Time Collaborations

By encouraging and making it easy for students to learn from each other, the lecturer will also find his teaching load reduced. The lecturer would be able to focus more on the higher-level concepts and principles than repeating and answering the tedious questions raised by each student again and again.

## Goal 4- Lecture Recording

Providing a second chance for the students to review the lecture is often helpful for students to strengthen their knowledge learned and discover new knowledge. Though there is worry that recording and sharing the video lecture online would compromise the student attendances, several recent studies [42, 43] report the favorable results that actually recording the lectures and sharing the archived lecture online would actually boost up the student attendances. The possible reasons are that students would find it more fruitful to come to the lecture first and re-study recorded lectures catch up the unabsorbed portion in classroom learning. For those students who have to miss one or two lectures, watching the recorded lectures will help them more easily to catch up with the next lecture in the class.

In our teaching environment, we have tested to record a lecture using the trial version Camtasia Studio. By simply starting a recording using Camtasia Studio when the lecture starts and stop it when the lecture finishes, this tool can record both the lecturer’s voice and the screen activity on the instructor’s PC into a compressed flash video of approximate 20 MB for a one-hour duration. The software is easy to start with and the process has least interference with other activities such as electronic inking. We recommend purchasing this Camtasia Studio software when new funding is available to support the faculties to try out recording and sharing the archived lectures online.

After supporting these 4 main pedagogical goals, our constructed Tablet PC based classroom promotes attractive new learning styles and likely receive positive feedbacks from students and lecturer. Currently, our teaching environment has attracted some Professors, who use ICIS Practicum lab for teaching, to try out the Tablet PC based teaching environment and software tools we have equipped for their daily teaching purposes. Generally, we receive positive perception on the environment we have setup. The software tools are generally not difficult to pick up. Initially, sometimes, the dual monitor setup can be messed up by a lecturer or some students because they are not familiar and do not understand how the dual monitor configuration works. These problems are often timely fixed by the technical staffs in ICIS Practicum lab.

# 3. Limitation and Future Extension

The pilot Tablet PC teaching environment is designed for classroom teaching only and it does not well support the distributed learning, i.e. students have to physically come to teaching environment in order to enjoy the teaching facilities. One of professors has enquired on whether it is possible to allow student attending the lecture through Internet. Since the Classroom Presenter by itself allows students to connect to the lecture's application remotely using Internet, the only problem would be the voice data are not transmitted. Our current workaround is to ask the lecturer to use communication tools, such as Skype, in a multicast mood to simultaneous transmit his voices to the remote students. We assume the voice synchronization and network latency problems are not too serious as the current broadband connection is usually fast enough. The current classroom setup can be extended to enable multimedia-rich distributed learning such as the joint collaboration courses with other universities. However, this will require additional funding to equip the overhead cameras and to connect the Classroom Presenter application on top of the Microsoft ConferenceXP platform. Additional studies are required to make sure smooth transmission of multimedia data.

**Title: Interactive Teaching and Learning at NTU**
**An introduction paragraph that briefly describes our project**
**Contents**

1. Online Signal Processing Lab
2. Teaching with Tablet Classroom
3. Active learning in the Virtual World
4. References
5. Related links

**Online Signal Processing Lab**

Paragraphs to elaborate the functionality + a representative picture + Examples + Link to our online Signal processing lab

**Teaching with Tablet Classroom**

Paragraphs on our Tablet teaching environment, the Classroom Presenter tool and how to use the tools + a representative picture + External links

**Active Learning in the Virtual World**

Paragraphs on the virtual learning environment + a representative picture + Link to our developed virtual learning world

**References**
**Related links**

Fig. 14 Brief Outline of the Wikipedia-Style Webpage for our Interactive Teaching and Learning Project

To publicize the entire TEF project, we have also considered building a Wikipedia-style website to integrate usage of different interactive technologies to enhance the teaching and learning. The proposed outline of the webpage is shown in Fig. 14. Additional work will be done on this if the there is future extension of the current development.

# 4. Conclusion

In this work, we have studied using Tablet PC technology to construct an interactive classroom teaching environment for higher education. Through literature review on the previous studies, we have summarized the main advantages of using pen-based computing technology in classroom teaching as well as the lessons learned from previous experiences. By taking all these into account, we have set up our Tablet PC learning environment to mainly support four pedagogical goals, including electronic inking anywhere, enhanced active learning and interactivity, easy student collaboration and lecture recording and archiving. Positive perception has been received from lecturers in the preliminary usage of the classroom. As one of the three parts in our TEF project, we have successfully completed in constructing a physical interactive learning environment.